%% file: main.tex
\pdfoutput=1
\documentclass[sigconf,natbib=true]{acmart}
\AtBeginDocument{%
  }

\usepackage{graphicx}
\usepackage{booktabs}

\copyrightyear{2026}
\acmYear{2026}
\setcopyright{cc}
\setcctype{by}
\acmConference[CIKM '26]{Proceedings of the 35th ACM International Conference on Information and Knowledge Management}{November 07--11, 2026}{Rome, Italy}
\acmBooktitle{Proceedings of the 35th ACM International Conference on Information and Knowledge Management (CIKM '26), November 07--11, 2026, Rome, Italy}
\acmDOI{10.1145/3799682.3840004}
\acmISBN{979-8-4007-2539-5/2026/11}
\begin{document}

\title{The Fellowship of the Query: Learning Retrieval Actions}

\author{Mohammed Al-Maamari}
\orcid{0000-0002-0127-8034}
\affiliation{%
  \institution{University of Passau}
  \city{Passau}
  \country{Germany}
}
\email{mohammed.al-maamari@uni-passau.de}

\author{Saber Zerhoudi}
\orcid{0000-0003-2259-0462}
\affiliation{%
  \institution{University of Passau}
  \city{Passau}
  \country{Germany}
}
\email{saber.zerhoudi@uni-passau.de}

\author{Michael Granitzer}
\orcid{0000-0003-3566-5507}
\affiliation{%
  \institution{University of Passau}
  \city{Passau}
  \country{Germany}
}
\email{michael.granitzer@uni-passau.de}

\author{Jelena Mitrovi\'{c}}
\orcid{0000-0003-3220-8749}
\affiliation{%
  \institution{University of Passau}
  \city{Passau}
  \country{Germany}
}
\email{jelena.mitrovic@uni-passau.de}

\renewcommand{\shortauthors}{Mohammed Al-Maamari, Saber Zerhoudi, Michael Granitzer, and Jelena Mitrovi\'{c}}

\begin{abstract}
Retrieval-augmented question answering requires control decisions about when to decompose a question, search, reformulate, extract evidence, synthesize facts, verify progress, and stop. We study whether trajectory fine-tuning can improve small language models (SLMs) as next-action controllers. We additionally evaluate a low-resource setting in which a single SLM serves as both the controller and the final-answer generator. From accepted teacher search traces, we build a seven-way action-prediction task, where the model predicts the next structured teacher action from the current trajectory state, and evaluate LoRA-supervised fine-tuning across SLMs and xSLMs as controllers. On 1,646 held-out action examples, Granite 4.1 3B trained on 13,194 actions reaches macro-F1 0.6536, compared with 0.1736 for zero-shot prompting of the same model and 0.5399 for a TF-IDF logistic-regression baseline. In an end-to-end controller/generator swap evaluation over 149 held-out trajectories, using the fine-tuned model for both roles improves Exact Match from 0.7530 to 0.7946 and token F1 from 0.7783 to 0.8295 compared with using the base model as both controller and generator. The cross-role conditions show that the fine-tuned controller increases evidence-fact recording when the generator is fixed, while controller-only final-answer gains are not statistically clear. Overall, trajectory supervision improves action prediction and evidence-recording behaviour in this evaluated pipeline. The code used in this paper is publicly available.\footnote{Code: \url{https://github.com/padas-lab-de/agent-action-controller}.}
\end{abstract}



\maketitle

\input{short_paper}

\end{document}

%% file: short_paper.tex
\section{Introduction}
\label{sec:introduction}

Retrieval-augmented generation (RAG) connects language models to external evidence sources for knowledge-intensive tasks \citep{Lewis:2020:NeurIPS,Mialon:2023:arXiv}. In multi-hop question answering, retrieval is not a single operation. A system must decide when to decompose a question, search, reformulate a query, extract evidence, synthesize intermediate facts, verify progress, and produce a final answer. This creates a search-control problem: final answer quality depends on both generation and the action sequence that constructs the evidence state.

Prompted agent frameworks such as ReAct combine reasoning and tool use \citep{yao2022react}. This can work when the model reliably follows an action protocol, but smaller models may repeatedly search, fail to extract grounded evidence, or fail to emit a valid finishing action. In this paper, we use \emph{xSLMs} for models below 1B parameters and \emph{SLMs} for models in the 1B--3B range. Our central question is whether trajectory supervision can improve an SLM as a next-action controller for search.

Prior work has studied interleaved retrieval and reasoning, retrieval self-reflection, query refinement, and search-agent training. IRCoT interleaves retrieval with chain-of-thought reasoning for multi-step QA \citep{trivedi2023interleaving}. Self-RAG trains models to retrieve, generate, and critique retrieved evidence \citep{asai2024self}. RQ-RAG focuses on learned query refinement actions such as rewriting, decomposition, and disambiguation \citep{chan2024rq}. Search-R1 trains models to interact with search engines over multi-turn retrieval trajectories using reinforcement learning \citep{jin2025search}. We study whether lightweight LoRA supervision enables xSLMs and SLMs to imitate a structured seven-action search-control protocol, and whether this changes evidence-fact recording in a constrained multi-hop retrieval pipeline. The accepted teacher trajectories lead to correct answers under our filtering protocol, but we do not assume that every action in those trajectories is optimal; another trajectory could reach the same answer with different or fewer actions.

The seven-action protocol separates recurring decisions in iterative information seeking: clarifying or decomposing the queries, issuing and reformulating queries, collecting evidence, combining evidence, checking progress, and completing the task. This motivation is consistent with prior models that describe search as an evolving and staged process rather than a single query-response event \citep{Bates:1989:Berrypicking,Kuhlthau:1991:ISP,Marchionini:2006:Exploratory}. Action-level imitation under this protocol evaluates whether the model chooses the correct next retrieval action.

We study this question using supervised trajectory fine-tuning. In our setting, \emph{action prediction} is the task of predicting the next structured action using as input the current question, trajectory state, and evidence context; the supervised target is the teacher action stored in an accepted trace. We evaluate this task across multiple SLMs and xSLMs, learning formulations, and training-set sizes. We then place the selected model in an end-to-end BM25 search pipeline \citep{Robertson:2009:FTIR} and evaluate final answers with SQuAD-style Exact Match and token F1 \citep{Rajpurkar:2016:arXiv}.

This work contributes a focused evaluation of search-flow control for RAG-style question answering as supervised action prediction from structured teacher trajectories. It studies trajectory fine-tuning for SLMs and xSLMs as controllers across selected models, training-set sizes, and multiclass versus binary action formulations \citep{granite2026,granite2025,qwen3.5,liu2026ministral,minicpm4,yu2025minicpmv45cookingefficient,yao2024minicpm,liquidai2025lfm2}, then evaluates the selected fine-tuned Granite 4.1 3B model in a repeated end-to-end controller/generator swap pipeline with paired bootstrap confidence intervals over the same held-out trajectories.

\section{Method}
\label{sec:method}

\subsection{Trajectory-to-Action Supervision}

The supervision source is a set of accepted search trajectories generated by a teacher rollout model, Qwen 3-Next 80B-A3B Instruct \citep{qwen3technicalreport}. 
Of 5,000 attempted rollouts, 1,490 trajectories passed the filtering pipeline and were used for training and evaluation. The rejected traces include 2,298 rejected by the gold-answer filter, 850 by deduplication, 254 by repeated-search filtering, and 108 by grounding checks. The resulting evaluation population is therefore conditional on teacher success under this filtering protocol; it is not an unfiltered sample of multi-hop questions. The accepted trajectories contain 16,488 action steps. Each step stores the teacher input messages, teacher JSON output, action tool, action parameters, observation, thought text, original query, dataset identifier, sample identifier, gold answer, and final citations when present.

Each trajectory is converted into supervised examples of the form $(q, s_t, e_t) \mapsto a_t$, where $q$ is the question, $s_t$ is the current structured state represented in the teacher prompt, $e_t$ is the evidence context visible in that prompt, and $a_t$ is the next action emitted by the teacher. The action space contains seven tools: \texttt{decompose}, \texttt{search}, \texttt{reformulate}, \texttt{extract}, \texttt{synthesize}, \texttt{verify}, and \texttt{finish}. For LoRA SFT, the model is trained to reproduce the teacher's full JSON action output; for classification methods, the label is the \texttt{action.tool} field. For an action at step $t$, the student input is the stored teacher prompt before that action. It contains the original question, prior state, previous tool history, and evidence preview visible to the teacher at that step. Other fields like the teacher thought text, gold answer, final answer, final citations, dataset identifier, and sample identifier are excluded from the action-prediction input. Gold answers are used for filtering and final-answer evaluation metadata, not as explicit input fields. The target output for generative controllers is JSON, but we only evaluate how accurately the SLM predicts the correct next action, we ignore the validity of the output JSON since we parse the predicted action.

Splitting is performed at both trajectory and originating-query level: no trajectory and no original query contributes action decisions to any other split to avoid any contamination. The split contains 1,192 train trajectories, 149 validation trajectories, and 149 test trajectories, corresponding to 13,194 train actions, 1,648 validation actions, and 1,646 test actions. The action distributions are similar across splits. For example, \texttt{search} is 40.1\%, 40.7\%, and 42.5\% in train, validation, and test, while \texttt{verify} is 1.2\%, 0.8\%, and 1.0\%. We report macro-F1 as the primary action-prediction metric, with accuracy and unparsed-output counts as secondary metrics. Unparsed cases are samples where robust action parsing cannot recover a predicted action; outputs are counted as parsed even when the JSON is invalid, as long as the action can be recovered.

\subsection{Learning Formats}

We evaluate zero-shot prompting, TF-IDF logistic regression, LoRA sequence classification, LoRA supervised fine-tuning (LoRA SFT), and one-vs-rest binary variants. LoRA adapters use rank $r=16$, $\alpha=32$, dropout 0.05, no bias terms, and attention projection targets corresponding to query, key, value, and output projections where available \citep{hu2022lora}. Target module names differ by architecture, but the common design is to adapt attention projections while keeping the base model fixed.

The LoRA SFT models are trained for two epochs with a learning rate of $2\cdot10^{-4}$, bfloat16, maximum training length 2,048 tokens, batch size 2, and gradient accumulation 4. Sequence-classification LoRA uses the same learning rate, epochs, precision, and maximum length, with batch size 4 and gradient accumulation 2. Both therefore use an effective batch size of 8 examples. We use the Hugging Face Trainer defaults for AdamW optimisation and a linear schedule with no warmup, evaluate and save once per epoch, keep one checkpoint, and load the best checkpoint by validation loss for LoRA SFT and validation macro-F1 for sequence classification. Generative evaluation uses greedy decoding with maximum input length 3,072 and maximum 512 new tokens. Invalid or unparsed generated actions are counted as unparsed rather than assigned to a valid class.

\subsection{End-to-End Pipeline}

The end-to-end agent has a controller role and a final-answer generator role. The same model can be used for both roles, or different models can be assigned to separate them. At each step, the controller emits a JSON action. Search actions query a deterministic BM25 retriever over a small per-question candidate corpus, not a global web or production-scale collection. For HotpotQA \citep{Yang:2018:EMNLP}, 2WikiMultiHopQA \citep{ho-etal-2020-constructing}, and MuSiQue \citep{trivedi-etal-2022-musique} examples, the corpus is built from original context paragraphs containing gold supporting paragraphs plus distractors. For FRAMES \citep{krishna2024factfetchreasonunified} examples, where such context paragraphs are not available, it uses the trace evidence pool. In the 149-trajectory test corpus, each question has 5 to 20 candidate documents, with mean 10.0 and median 10. Gold-supporting documents are present for mapped source-dataset examples, with mean 2.4 gold documents per question. The controller does not search outside this per-question corpus.

Extraction actions record evidence facts by matching requested facts against retrieved evidence. A fact is counted as evidence-matched if it has a case-insensitive substring match in retrieved evidence, or a token-overlap match of at least 0.6. This is an automatic evidence-linking heuristic, not an independent human or entailment-based factuality audit. After the controller finishes or reaches the step budget, the final-answer generator receives the accumulated state and retrieved evidence through a separated answer prompt.

The maximum controller budget is 12 steps. If the controller does not finish within this budget, the system still calls the final-answer generator from the current accumulated state. Thus, the reported answer quality reflects the separated final-answer prompt rather than a learned early-stopping policy.

\section{Experiments}
\label{sec:experiments}

We organize the experiments around three questions. RQ1 asks whether trajectory fine-tuning improves action prediction for the evaluated SLMs and xSLMs. RQ2 asks how the result changes with training-set size and with one multiclass controller versus one binary controller per action. RQ3 asks whether action-prediction gains are reflected in the constrained end-to-end search pipeline. Model and configuration choices were selected using the validation split; all numbers reported below are test-set results unless explicitly described as validation results.

\subsection{RQ1: Action Prediction Across Models}

Table~\ref{tab:action} reports representative 7-way action-prediction results on the same 1,646-example test split. For Granite 4.1 3B, LoRA SFT improves macro-F1 over zero-shot prompting of the same model. Granite 4.1 3B trained on the full 13,194-sample training split reaches test macro-F1 0.6536. Ministral 3B with LoRA sequence classification gives the highest accuracy among the reported results, but lower macro-F1 than the best LoRA SFT results.

\begin{table}[t]
\centering
\small
\caption{Action-prediction results on 1,646 held-out examples. Unp. refers to unparseable.}
\vspace{-0.5em}
\label{tab:action}
\setlength{\tabcolsep}{3pt}
\begin{tabular}{@{}p{0.30\linewidth}p{0.22\linewidth}ccc@{}}
\toprule
Model & Method & Acc. & Macro-F1 & Unp. \\
\midrule
Granite 4.1 3B & LoRA SFT & 0.7581 & 0.6536 & 1 \\
Ministral 3B & LoRA SFT & 0.7497 & 0.6443 & 0 \\
Qwen 3.5 2B & LoRA SFT & 0.7448 & 0.6414 & 0 \\
Granite 4.0 1B & LoRA SFT & 0.7418 & 0.6391 & 1 \\
Qwen 3.5 0.8B & LoRA SFT & 0.7179 & 0.6060 & 1 \\
Granite 4.0 350M & LoRA SFT & 0.6586 & 0.5683 & 0 \\
TF-IDF logreg & Baseline & 0.7102 & 0.5399 & 0 \\
Ministral 3B & LoRA seqcls & 0.7861 & 0.6330 & 0 \\
Granite 4.1 3B & Zero-shot & 0.4399 & 0.1736 & 0 \\
\bottomrule
\end{tabular}
\end{table}

The model comparison is descriptive. Granite 4.0 350M reaches macro-F1 0.5683, Granite 4.0 1B reaches 0.6391, and Granite 4.1 3B reaches 0.6536 when trained on the full action split. We use Granite 4.1 for the 3B model because Granite 4.1 does not have 1B or 350M variants. Because the 3B result is therefore also from a newer Granite generation, this comparison should not be interpreted as a controlled parameter-scaling result. The Qwen comparison shows a similar direction, with Qwen 3.5 0.8B at macro-F1 0.6060 and Qwen 3.5 2B at 0.6414 under LoRA SFT.

The learning format also matters. In the 3B-class comparison, LoRA sequence classification gives higher accuracy for Granite and Ministral, while LoRA SFT gives higher macro-F1. We do not treat this as a general method ranking. The \texttt{verify} class remains difficult in all settings, with only 17 held-out examples.

\subsection{RQ2: Data Size and Action Formulation}

\begin{figure}[t]
  \centering
  \includegraphics[width=.82\linewidth]{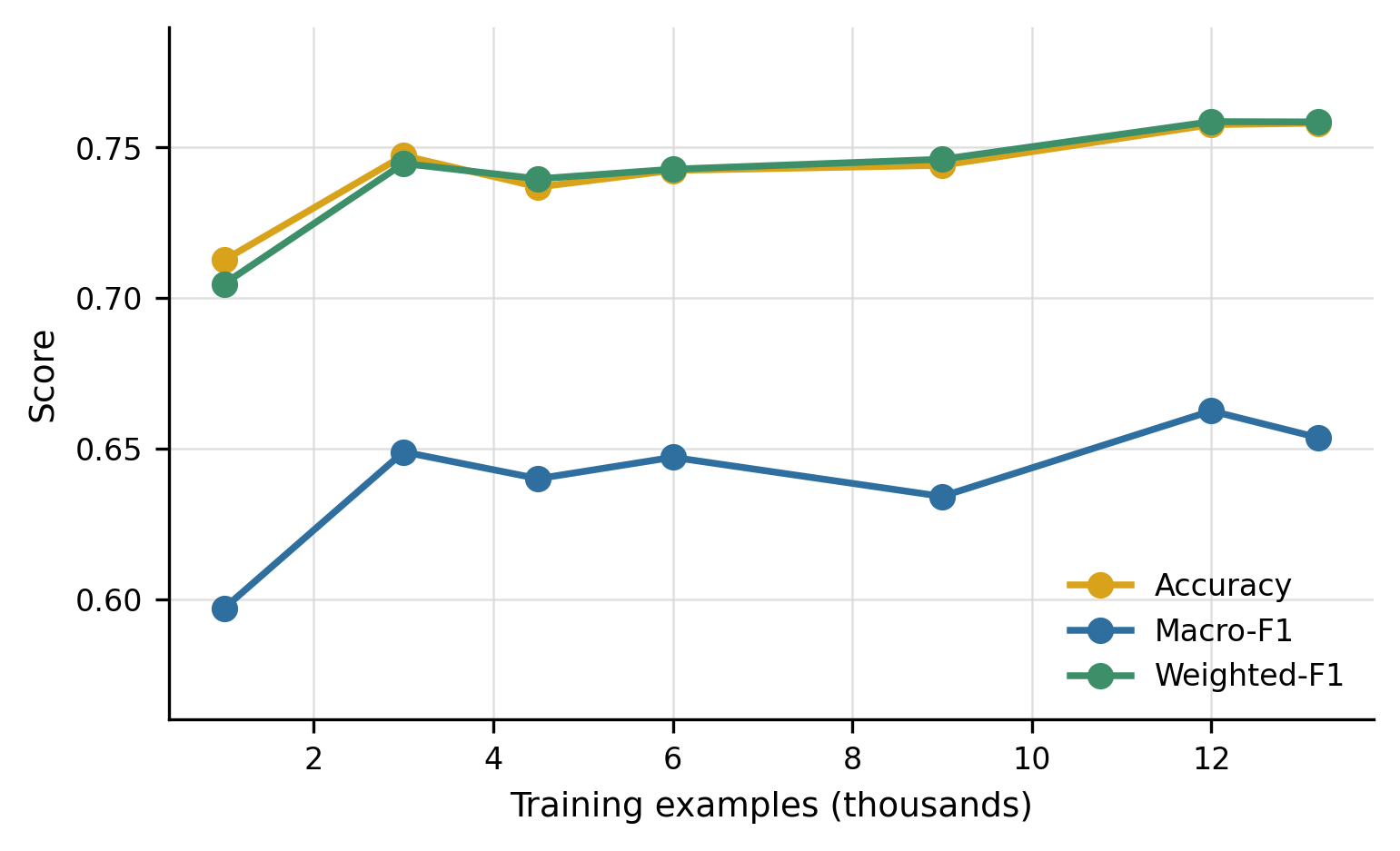}
  \caption{Granite 4.1 3B LoRA-SFT scaling as the number of training samples increases.}
  \label{fig:granite-sft-training-size-metrics}
\end{figure}

Figure~\ref{fig:granite-sft-training-size-metrics} shows the training-size sweep for Granite 4.1 3B with LoRA SFT. In the reported sweep, the 3,000-training-sample setting obtains macro-F1 0.6489, close to the best observed value of 0.6625 at 12,000 training samples and to 0.6536 with the full 13,194-sample training split. This suggests that, for this action protocol and accepted-trace distribution, a few thousand supervised trajectory steps can already provide much of the observed action-prediction gain for this SLM. Accuracy and weighted-F1 continue to increase slightly at larger sizes, while macro-F1 peaks at 12,000 training samples and is similar on the full training set. For consistency with the final full-data model, the main Granite LoRA SFT result is reported as macro-F1 0.6536 when trained on all 13,194 training actions. Because the split is class-imbalanced, small changes in rare classes can move macro-F1 without changing overall accuracy much.

We next evaluate the second RQ2 formulation choice. The default setup treats action prediction as a single seven-class decision over the seven tools; we compare this multiclass controller against one binary LoRA SFT controller per action using Qwen 3.5 0.8B. Table~\ref{tab:binary} summarizes the aggregate results. Accuracy is not reported for the independent binary controllers because they are evaluated one-vs-rest and do not by themselves produce a single action label per sample. The single multiclass controller gives higher macro-F1 and higher macro-recall. The independent binary controllers give higher macro-precision, but they lose recall on several low-support or confusable classes. A naive ensemble that selects the action with the highest binary ``yes'' score performs worse than the multiclass controller.

\begin{table}[t]
\centering
\small
\caption{Multiclass versus binary action formulations with Qwen 3.5 0.8B LoRA SFT.}
\vspace{-0.5em}
\label{tab:binary}
\setlength{\tabcolsep}{3pt}
\begin{tabular}{@{}p{0.36\linewidth}cccc@{}}
\toprule
Formulation & Acc. & Macro-P & Macro-R & Macro-F1 \\
\midrule
Single multiclass controller & 0.7005 & 0.6196 & 0.5979 & 0.6021 \\
Seven binary controllers & n/a & 0.6722 & 0.5561 & 0.5680 \\
Naive binary ensemble & 0.5365 & 0.4277 & 0.4048 & 0.3978 \\
\bottomrule
\end{tabular}
\end{table}
Direct classification performed worse than SFT: The binary formulation has higher macro-precision but lower macro-recall and macro-F1. The weak naive ensemble indicates that direct comparison of independently trained binary scores is ineffective in this setting. For the end-to-end pipeline, we therefore use a single multiclass LoRA SFT controller.

\subsection{RQ3: End-to-End Search}

We evaluate the selected Granite 4.1 3B models in an end-to-end controller/generator swap pipeline. The controller builds the search state through structured actions, and the final-answer generator is then called with the same separated answer prompt over the accumulated state and retrieved evidence. This design evaluates the base and fine-tuned models in four role combinations: base/base, fine-tuned/base, base/fine-tuned, and fine-tuned/fine-tuned. The base/base and fine-tuned/fine-tuned conditions correspond to using one model for both roles, while the cross-role conditions isolate the controller and generator roles. Each condition was run five times with different random seeds on the same 149 held-out accepted test trajectories, using stochastic decoding with temperature 0.7, the same BM25 retrieval environment, and \texttt{MAX\_STEPS=12}.

We evaluate the final answer with SQuAD-style Exact Match and token F1 against the gold answer and aliases. These metrics measure whether the completed pipeline produces the expected answer string. To describe the controller's intermediate behaviour, we also report evidence-fact rate: the fraction of trajectories in which at least one extracted intermediate fact is automatically matched to retrieved evidence. This diagnostic measures whether the controller records evidence-grounded intermediate facts before final answer generation. It is based on the extraction heuristic described in Section~\ref{sec:method}, and should not be interpreted as a human or entailment-based factuality judgment.

\begin{table}[t]
\centering
\small
\caption{End-to-end controller/generator swap evaluation.}
\vspace{-0.5em}
\label{tab:e2e}
\setlength{\tabcolsep}{3pt}
\begin{tabular}{@{}p{0.25\linewidth}p{0.25\linewidth}ccc@{}}
\toprule
Controller & Generator & EM & F1 & Ev.-fact rate \\
\midrule
Base & Base & 0.7530 & 0.7783 & 0.0121 \\
Fine-tuned & Base & 0.7664 & 0.8044 & 0.6161 \\
Base & Fine-tuned & 0.7691 & 0.7997 & 0.0134 \\
Fine-tuned & Fine-tuned & 0.7946 & 0.8295 & 0.6389 \\
\bottomrule
\end{tabular}
\end{table}

Table~\ref{tab:e2e} shows that using the fine-tuned model for both roles improves answer quality relative to base/base in this separated-prompt pipeline: EM increases from 0.7530 to 0.7946, and token F1 increases from 0.7783 to 0.8295. The paired bootstrap interval for the combined token-F1 gain is [0.0021, 0.1040], while the Exact Match interval crosses zero. The cross-role conditions show that the fine-tuned controller has the clearest effect on evidence recording. With the base generator fixed, replacing the base controller with the fine-tuned controller increases the evidence-fact rate by 0.60 and the mean number of evidence facts by 2.25. With the fine-tuned generator fixed, the same controller replacement increases the evidence-fact rate by 0.62 and the mean number of evidence facts by 2.28. The corresponding controller-only answer-quality deltas are smaller and have bootstrap intervals crossing zero. Thus, the end-to-end result supports a controller effect on evidence-fact recording and a modest combined improvement when both roles use the fine-tuned model, but it does not establish a statistically clear controller-only improvement in final-answer quality.

The 149 test trajectories contain 105 2WikiMultiHopQA, 41 HotpotQA, 2 FRAMES, and 1 MuSiQue examples. FRAMES and MuSiQue have too few examples for dataset-level conclusions.

\section{Discussion}
\label{sec:discussion}

The experiments support three main observations. First, action prediction is learnable for this seven-action protocol. Granite 4.1 3B trained on the full 13,194-sample training split improves substantially over zero-shot prompting of the same backbone, and other evaluated LoRA-SFT models obtain macro-F1 between 0.5683 and 0.6443. Second, the task is affected by class imbalance. Macro-F1, accuracy, and weighted-F1 do not always select the same method, and the rare \texttt{verify} class remains unreliable. Third, the end-to-end controller/generator swap shows that the trained controller increases evidence-matched intermediate facts under the automatic heuristic when the generator is held fixed. The answer-quality changes from replacing only the controller are modest and have bootstrap intervals crossing zero, while using the fine-tuned model for both roles gives a small token-F1 gain over base/base.

We also evaluate the base and fine-tuned Granite 4.1 3B models on three external multiple-choice benchmarks as a general-capability retention check. Under the same \texttt{lm-evaluation-harness} protocol \citep{biderman2024lessons}, fine-tuning reduces MMLU accuracy \citep{hendrycks2021mmlu} from 0.6554 to 0.5520, HellaSwag \citep{zellers-etal-2019-hellaswag} from 0.7911 to 0.6852, and ARC-Challenge \citep{clark2018arc} from 0.6502 to 0.5589, indicating a measurable trade-off on these non-trajectory tasks. Because training is adapter-based, the controller adapter can be used for search-flow control while the base model remains available for generation; in the end-to-end swap, however, using the fine-tuned model for both controller and generator still gives the strongest answer scores among the evaluated role combinations.

The results should be interpreted with several constraints. The evaluation set is acceptance-filtered: both action prediction and end-to-end QA are measured on questions for which the teacher produced an accepted trajectory. Action prediction measures agreement with teacher-labeled targets, not proof that the predicted action is uniquely correct. The retrieval environment is a per-question candidate corpus with gold supporting paragraphs and distractors where available, not an open retrieval collection. The end-to-end evaluation uses a separated final-answer prompt, so its answer scores characterize this specific pipeline rather than all possible agent finalization interfaces. The evidence-fact metric is based on an automatic matching heuristic and should not be read as human-verified factuality.

\section{Conclusion}
\label{sec:conclusion}

We evaluated trajectory fine-tuning for search-control actions in SLMs and xSLMs. On an acceptance-filtered held-out 7-way action-prediction task, Granite 4.1 3B LoRA SFT trained on the full 13,194-sample training split reaches macro-F1 0.6536, improving over zero-shot prompting and a TF-IDF baseline. In the reported training-size sweep, the 3,000-training-sample setting obtains macro-F1 0.6489, close to the best observed value of 0.6625, and a single multiclass controller is stronger in aggregate than independent binary action controllers for the evaluated Qwen 3.5 0.8B setting. In the end-to-end controller/generator swap, using the fine-tuned Granite 4.1 3B model for both roles improves token F1 relative to base/base, and replacing only the controller substantially increases evidence-fact recording under a fixed generator. The evidence supports a focused conclusion: trajectory supervision can improve action prediction and evidence-fact recording behaviour for an SLM in this evaluated pipeline, while unfiltered-query evaluation and robust final-answer generation remain open issues.

\section{GenAI Usage Disclosure}

The authors used GenAI to edit and improve text written for this paper. All AI-generated text was carefully reviewed prior to submission. In addition, GenAI-assisted code support tools were employed during development. Specific implementation-related questions were also posed to chat-based GenAI models, and the responses were used as a basis for adapting the code accordingly.

\begin{acks}
This work was funded by the Bavarian State Ministry of Economic Affairs, Regional
Development, and Energy (StMWi) under the funding code VAL-2505-0003, and by the
project \emph{FrAInderl -- Allianz f\"ur generative KI im Bayerischen Wald},
co-funded by the European Union (European Regional Development Fund, ERDF) and the
Bavarian State Ministry of Science and the Arts (StMWK).
\end{acks}

\bibliographystyle{ACM-Reference-Format}
\balance
\bibliography{sample-base}